\documentclass{article}

\newcommand{\documentdate}{19 01 2014}
\usepackage{a4wide,graphicx,color}
\usepackage{amsmath,amsfonts}
\usepackage{geometry}

\newtheorem{theorem}{Theorem}

\newtheorem{lemma}{Lemma}

\newtheorem{problem}{Problem}
\newtheorem{proposition}{Proposition}

\newenvironment{proof}[1][Proof]{\noindent\textbf{#1.} }{\ \rule{0.5em}{0.5em}}
\input{tcilatex}
\title{On Convergence in the Spatial AK Growth Models}
\author{G. Aldashev, S. Aldashev, T. Carletti}
\date{\documentdate}

\begin{document}
\begin{titlepage}

\includegraphics[height=3.5cm]{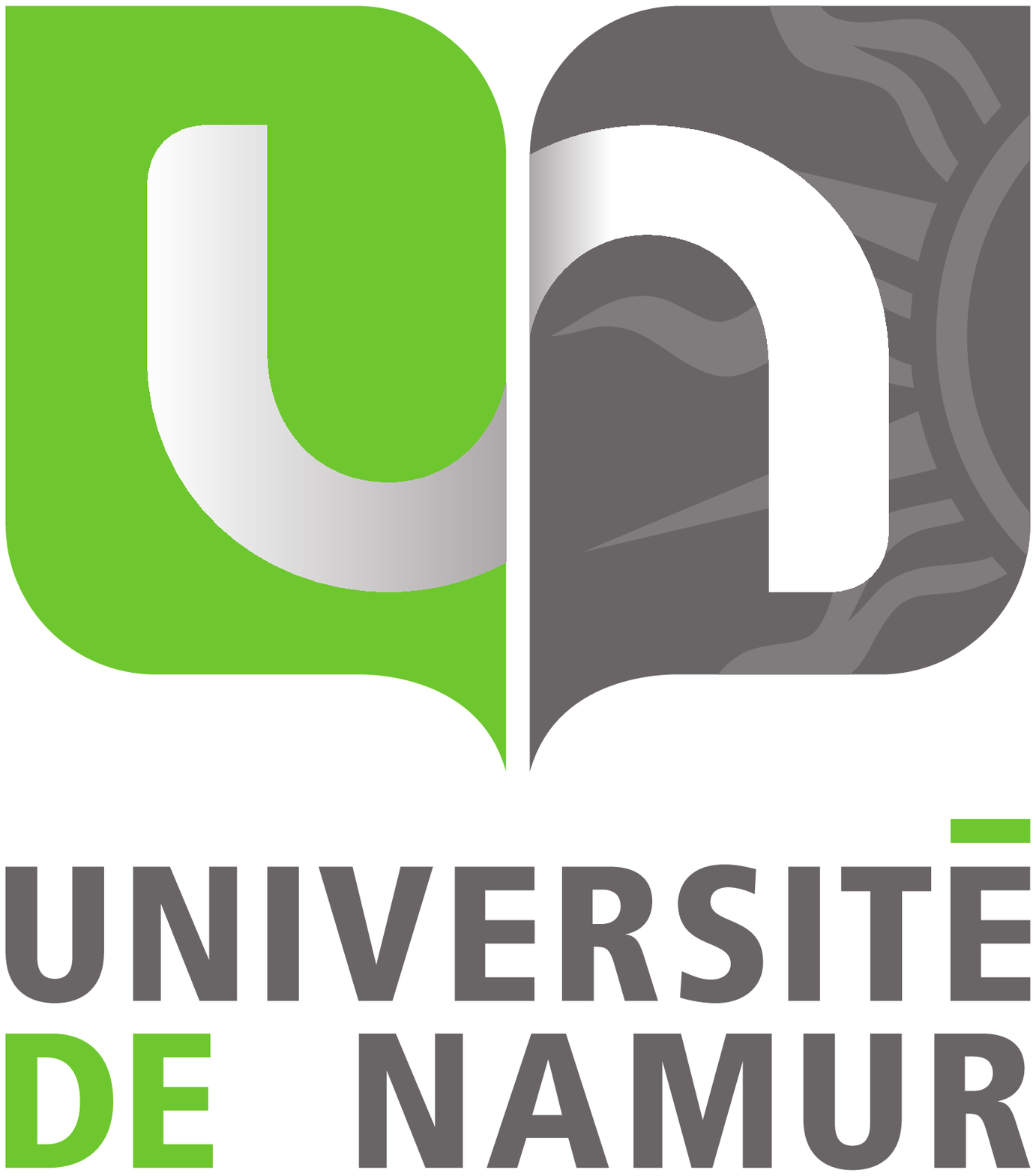}\hfill\includegraphics[height=2.5cm]{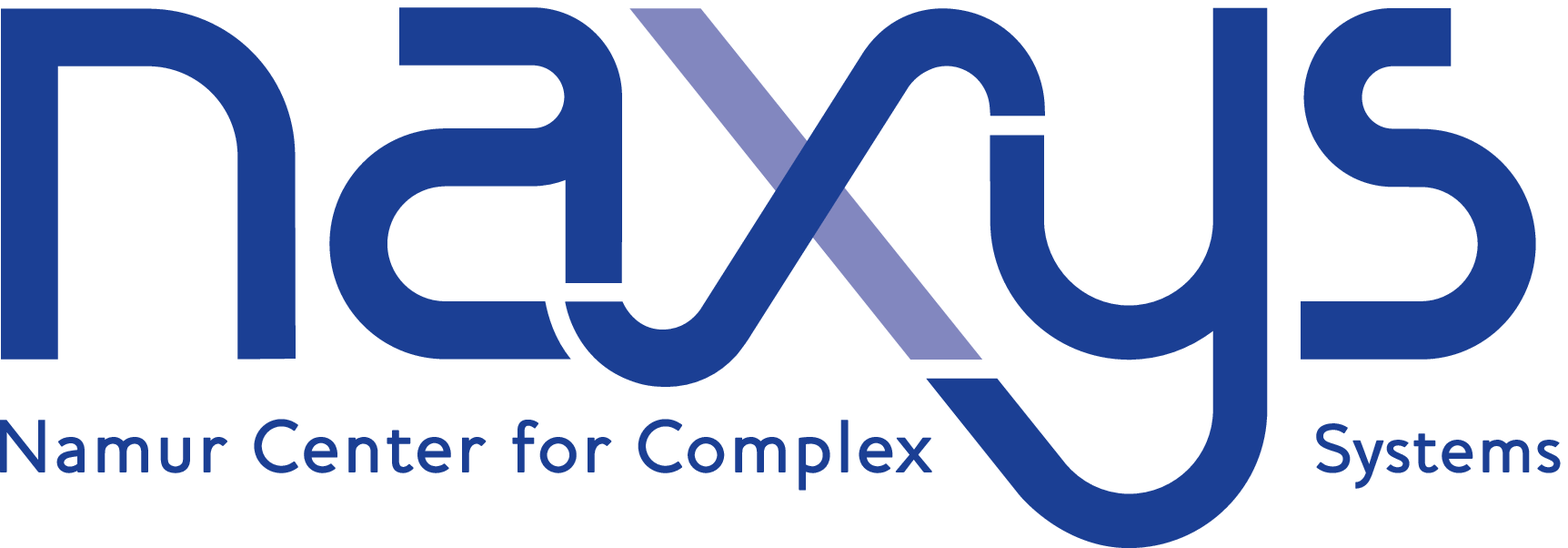}

\vspace*{2cm}
\hspace*{1.3cm}
\fbox{\rule[-3cm]{0cm}{6cm}\begin{minipage}[c]{12cm}
\begin{center}
\vspace{1cm}
On Convergence in the Spatial AK Growth Models\\
\vspace{1cm}
by G. Aldashev, S. Aldashev, T. Carletti \\
\mbox{}\\
Report naXys-03-2014 \hspace*{20mm} \documentdate \\
\vspace{2cm}
\includegraphics[height=7cm]{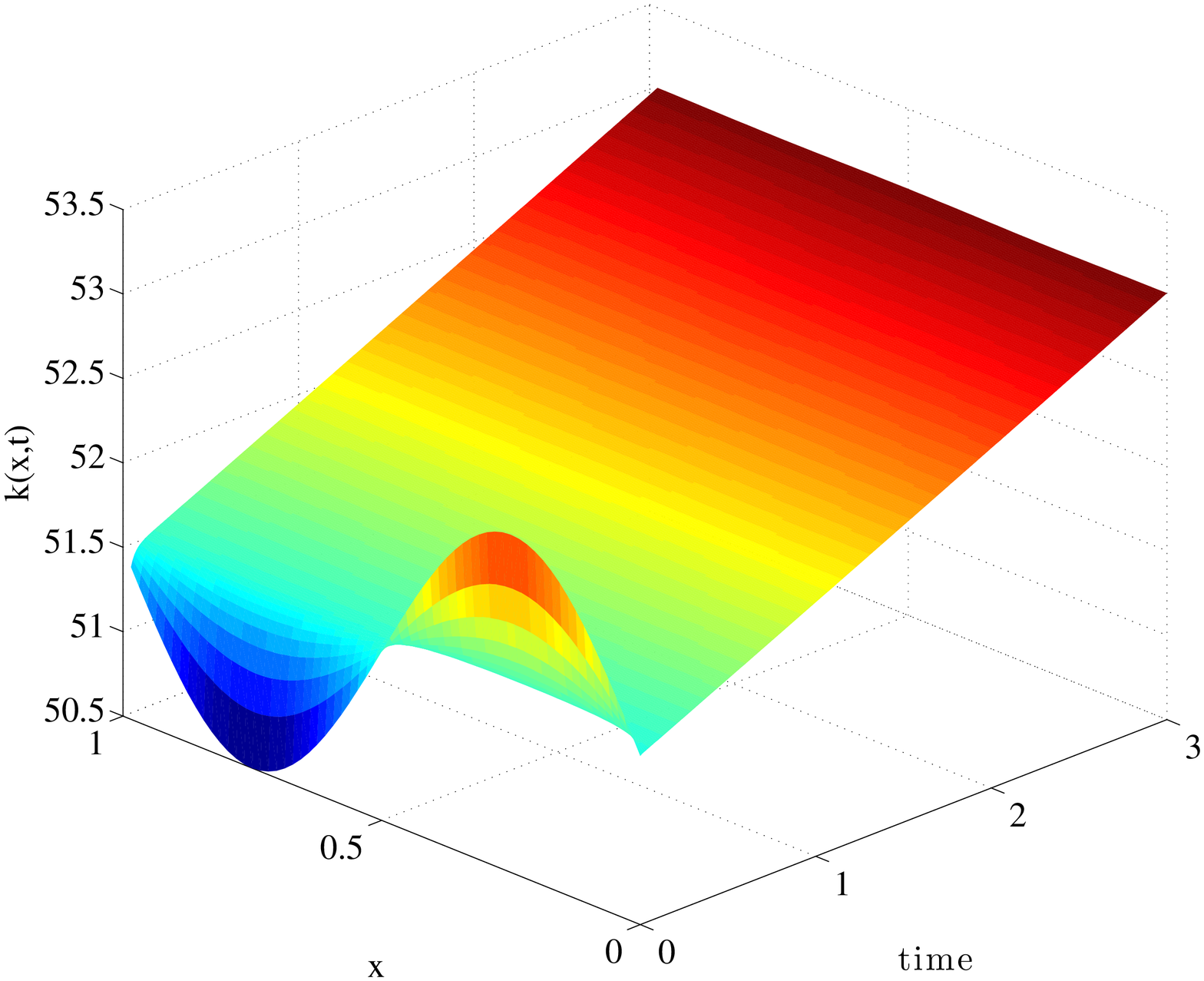}
\mbox{}
\end{center}
\end{minipage}
}

\vspace{2cm}
\begin{center}
{\Large \bf Namur Center for Complex Systems}

{\large
University of Namur\\
8, rempart de la vierge, B5000 Namur (Belgium)\\*[2ex]
{\tt http://www.naxys.be}}

\end{center}

\end{titlepage}

\newpage
\title{On Convergence in the Spatial AK Growth Models\thanks{%
This research has been partially financed by the Belgian National Bank. The
second author kindly acknowledges the support of the Namur Center for
Complex Systems (naXys) visiting project. This paper presents research
results of the Belgian Network DYSCO (Dynamical Systems, Control, and
Optimization), funded by the Interuniversity Attraction Poles Programme,
initiated by the Belgian State, Science Policy Office. The scientific
responsibility rests with its author(s).}}
\author{Gani Aldashev\thanks{%
Corresponding author. Department of Economics and CRED, University of Namur.
Mailing address: Department of Economics, 8 Rempart de la Vierge, 5000
Namur, Belgium. Email: gani.aldashev@unamur.be. Tel.: +3281724862, fax:
+3281724840.} \and Serik Aldashev\thanks{%
Abai Kazakh National University. Mailing address: Faculty of Mathematics and
Physics, Tole Bi Street, 86, 050000 Almaty, Kazakhstan. Email:
aldash51@mail.ru. } \and Timoteo Carletti\thanks{%
NaXyS and Department of Mathematics, University of Namur. Mailing address:
Rempart de la Vierge 8, 5000\ Namur, Belgium. Email:
timoteo.carletti@unamur.be. }}
\maketitle

\begin{abstract}
Recent research in economic theory attempts to study optimal economic growth
and spatial location of economic activity in a unified framework. So far,
the key result of this literature - asymptotic convergence, even in the
absence of decreasing returns to capital - relies on specific assumptions
about the objective of the social planner. We show that this result does not
depend on such restrictive assumptions and obtains for a broader class of
objective functions. We also generalize this finding, allowing for the
time-varying technology parameter, and provide an explicit solution for the
dynamics of spatial distribution of the capital stock.\bigskip


\newpage
\end{abstract}

\setlength{\baselineskip}{20pt} \newpage 

\section{Introduction}

Two major fields of research in theoretical economics - theory of economic
growth and models of new economic geography - have been developing
separately and independently until recently. The need to integrate them in a
common framework, as well as the underlying fundamental mechanisms between
endogenous growth and agglomeration of economic activity - in particular,
increasing returns to scale - was forcefully stated by Lucas (1988) and, in
more detail, by Krugman (1997). Over the last decade, economists have been
trying to construct such unified analytical frameworks that allow to capture
and describe the evolution of economic activity both in space and in time.
Most of this recent literature has been summarized in a survey by Desmet and
Rossi-Hansberg (2010), as well as in Chapters 18\ and 19\ of Acemoglu
(2008). What emerges from these surveys is that while a commonly agreed
framework has not yet emerged, research has been evolving using several
different and highly promising approaches.

One of these approaches is to build optimal growth models with spatial
dimension allowing for continuous space and time, in which spatio-temporal
evolution of capital stock is described by a partial differential equation,
usually of the diffusion (parabolic) type. This framework, initially
suggested by Isard and Liossatos (1979), has been rigorously developed
recently by Brito (2004) and Boucekkine et al. (2009), for the spatial
version of Ramsey-Cass-Koopmans optimal growth model (i.e. with decreasing
returns to capital) and by Boucekkine et al. (2013) for an AK growth model.

The major result in all of these papers is the asymptotic disappearance of
spatial inequality, i.e. convergence of the capital stock over time to the
same level in all the regions, despite the initially heterogeneous spatial
distribution of capital. This is not too surprising for the spatial
Ramsey-Cass-Koopmans model, given the decreasing returns to capital, and
thus an intuitive economic reason for the flow of capital from more
capital-abundant to capital-scarce regions.\ However, the convergence result
is much more surprising for the spatial AK endogenous growth model, given
that the returns to capital are constant; it is well known that the
non-spatial versions of the AK model exhibit non-convergence (Romer 1986,
Lucas 1988, Rebelo 1991).

Boucekkine et al. (2013) rely on a specific assumption on the objective of
the benevolent social planner: she treats equally all the individuals,
independently of their location (and thus initial endowment of capital).\
Given this, the social planner chooses to smooth the (detrended)
consumption, both across space and time. This assumption could be difficult
to justify according to certain ethical criteria, for example, giving more
weight to individuals located in areas with a lower initial level of capital
stock. Thus, a natural question arises: How general is the convergence
result in the spatial AK model? In particular, does this surprising result
rely on the particular assumptions concerning the objective function of the
benevolent social planner?

This note provides an answer to this question and generalizes the findings
of Boucekkine et al. (2013) in two important ways. First, we find that the
convergence result does not rely on restrictive assumptions on the objective
of the social planner.\ In particular, we show that the asymptotic
convergence result holds for a program of the social planner with any
objective function which gives rise to a continuous consumption function,
provided that the present discounted value of the flow of maximum
consumption (on the entire space) does not exceed the initial capital stock
at any point of the space.

Second, we show that these results hold for the AK production function in
which technology ($A$) evolves over time. This is an important
generalization, because technological change can potentially alter the
spatial dynamics of capital stock, and thus a priori it is unclear whether
the convergence result would hold in such settings.

The rest of the paper is organized as follows. Section 2 presents the
model.\ Section 3 develops our analytical results.\ Section 4 highlights
directions for future research and concludes. Technical proofs are relegated
to the Appendix.

\section{Model}

In modelling the spatial economic growth process, we follow the approach
initiated by Brito (2004) and further developed by Boucekkine et al. (2009)
and Boucekkine et al. (2013).

Spatial dimension is modelled as a circle of radius one, on which atomistic
economic agents are assumed to be uniformly distributed. Thus, letting $x$
to denote the geographic location of an agent, we have $x\in \mathbb{T}%
=[0,2\pi ]$, where boundary points of $\mathbb{T}$ coincide. This circle is
a stylized representation of different regions in a country (in case of a
closed economy), or - allowing for perfect capital mobility - can represent
the global economy. Time is assumed to be continuous and evolves from zero
to infinity: $t\in \mathbb{R}^{+}=[0,\infty )$.

Our main object of interest is the spatial distribution of capital stock and
its evolution over time. We denote the capital accumulated at moment $t$ in
point $x$ with $k(x,t)$. The initial distribution of capital stock $k(x,0)$
on the circle is a known function $k_{0}(x)$.

The production function is of AK-type (see Acemoglu, 2008, chapter 11, for
detailed analysis of the properties of the non-spatial growth models with
the AK aggregate production function), i.e. the returns to capital are
constant.\ Importantly, we allow for technology to change over time
(although the technological frontier in any given moment is the same in
every point of the circle). Thus, $A(x,t)=A(t)$ denotes the level of
technology at time $t$ in the whole space.

The instantaneous budget constraint of the agent located at point $x$ at
moment $t$ is%
\begin{equation}
y(x,t)=c(x,t)+\tau (x,t)+s(x,t),  \label{budget}
\end{equation}%
and simply states that the production $y(x,t)$ is divided between
consumption $c(x,t)$, trade balance $\tau (x,t)$ (given that each region is
a small open economy), and saving $s(x,t)$. Saving represents capital
accumulation into the next instant of time:%
\begin{equation*}
s(x,t)=k_{t}(x,t).
\end{equation*}%
As stated above, production uses the linear AK-technology:%
\begin{equation*}
y(x,t)=A(t)k(x,t).
\end{equation*}%
Finally, concerning the trade balance, we assume that there is perfect
capital mobility.\ In other words, consider a region (arc) $R$ of the
circle. The consumption in excess of unsaved output minus consumption of
"domestic" output (i.e. of output produced within the region) comes from
other regions, which is reflected by the trade balance of this region.
However, in the balance of payments of the region $R$, this excess
consumption has to be financed by capital outflows.\ Thus, the regional
trade balance simply equals the symmetric of the inflow of capital from one
of its border minus the outflow from the other border:%
\begin{equation}
\int_{R}\tau (x,t)dx=-[k_{x}(b,t)-k_{x}(a,t)],  \label{tradebalance}
\end{equation}%
where $b$ and $a$ are the boundaries of the region $R$. Using the
Fundamental Theorem of Calculus, the trade balance can thus be written as 
\begin{equation*}
\int_{R}\tau (x,t)dx=-\int_{R}k_{xx}(x,t)dx,
\end{equation*}%
which, for a length of the region $R$ tending to zero, simply becomes%
\begin{equation*}
\tau (x,t)=-k_{xx}(x,t).
\end{equation*}

Thus, the instantaneous budget constraint (\ref{budget}) can be written as
the following equation of motion of capital:%
\begin{equation}
A(t)k(x,t)=c(x,t)-k_{xx}(x,t)+k_{t}(x,t),  \label{eqmotion}
\end{equation}%
and this constraint must hold for any point $x$ and moment $t$.

Moreover, given that we represent the space as a circle, the values of the
capital stock must coincide at the endpoints of the interval $\mathbb{T}%
=[0,2\pi ]$, and the smooth-pasting condition must also hold, at any moment $%
t$:%
\begin{equation}
k(0,t)=k(2\pi ,t)\text{ and }k_{x}(0,t)=k_{x}(2\pi ,t)\text{.}
\label{endpoints}
\end{equation}

The problem of optimal growth in this economy is that of a social planner
that maximizes a certain objective function $J(k_{0},c(x,t))$ by choosing
the consumption function $c(x,t)$, subject to the instantaneous budget
constraint (\ref{eqmotion}), the boundary value conditions (\ref{endpoints}%
), and the initial value condition $k(x,0)=k_{0}(x)$. Clearly, the value of
the capital stock $k(x,t)$ must be non-negative everywhere and in any moment
of time. Formally, this accounts to:

\begin{problem}
\label{Prob1} Find a \emph{non-negative classical solution}, namely a
continuous function in the closed domain $\bar{\Omega}$, where ${\Omega }=%
\mathbb{T}\times \mathbb{R}^{+}$, twice-continuously differentiable with
respect to $x$ in $\Omega $, of the linear parabolic partial differential
equation 
\begin{equation}
k_{t}=k_{xx}+A(t)k(x,t)-c(x,t)\quad \forall (x,t)\in \Omega \,,
\label{eq:PDEgen}
\end{equation}%
that satisfies the \emph{initial condition} 
\begin{equation}
k(x,0)=k_{0}(x),\quad \forall x\in \mathbb{T}\,.  \label{eq:IVC}
\end{equation}
\end{problem}

The problem of the social planner is a highly complicated
infinite-dimensional optimal control problem, where complications
essentially arise because one of the constraints is in the form of a partial
differential equation. In a key contribution, Boucekkine et al. (2013)
develop an analytical methodology that allows to overcome this challenge by
adapting the dynamic programming methods to this infinite-dimensional
problem.\ However, they need to impose a specific form on the objective
function of the social planner, in order to obtain a characterization of the
optimal consumption function $c(x,t)$.

Instead, we attack this problem differently.\ We study the problem of
finding a non-negative classical solution of the partial differential
equation describing the equation of motion of capital stock, for a general
(continuous) consumption function $c(x,t)$. In doing so, we determine two
different sufficient conditions on the consumption function that guarantee
the uniqueness and non-negativity of the explicit solution of the PDE
problem. The first condition leads to a space-invariant consumption function
(and is thus equivalent to the one posited by the objective function of the
social planner in Boucekkine et al. 2013).\ However, the second condition is
more general, and has a different economic interpretation. Next, we show
that the asymptotic properties of the solution are similar to the ones
determined by Boucekkine et al. (2013); in particular, we prove the
convergence of the capital stock in every point of the circle to the same
level as $t\rightarrow \infty $. Crucially, the second sufficient condition
on the consumption function that we find is considerably weaker than those
of Boucekkine et al. (2013).\ We thus show that the convergence result in
the spatial AK model is not driven by a particular objective function of the
social planner.

\section{Analysis}

The aim of this section is twofold: (i) to provide solutions (Theorems~\ref%
{thm:main1} and~\ref{thm:main}) of Problem~\ref{Prob1} under two different
assumptions on the consumption function $c(x,t)$, and (ii) to study the
asymptotic behavior of such solutions (Propositions~\ref{prop:ktlarge} and~%
\ref{prop:ktlarge1}).

Assuming the spatio-temporal consumption function to be a smooth concave
function with respect to the spatial variable for any positive time moment,
the following result holds:

\begin{theorem}
\label{thm:main1} Let $\Omega =\mathbb{T}\times \mathbb{R}^{+}$. Assume the
functions $A:\mathbb{R}^{+}\rightarrow \mathbb{R}^{+}$, $k_{0}:\mathbb{T}%
\rightarrow \mathbb{R}^{+}\cup \{0\}$ and $c:\mathbb{T}\times \mathbb{R}%
^{+}\rightarrow \mathbb{R}^{+}\cup \{0\}$ are continuous in their respective
domains. Assume also that 
\begin{equation}
c_{xx}(x,t)\leq 0\quad \forall (x,t)\in \mathbb{T}\times \mathbb{R}^{+}\,,
\label{eq:cconc}
\end{equation}%
and 
\begin{equation}
k_{0}(x)\geq \int_{0}^{t}e^{-\int_{0}^{s}A(z)\,dz}c(x,s)\,ds\quad \forall
(x,t)\in \Omega \,.  \label{eq:Ak0c1}
\end{equation}

Then the Problem~\ref{Prob1} admits a unique non-negative classical solution.
\end{theorem}

Note that given the coincidence of the endpoints of the interval $\mathbb{T}%
=[0,2\pi ]$ and the smooth-pasting condition (\ref{endpoints}), the
concavity assumption (\ref{eq:cconc}) leads to a consumption function that
is invariant in space. Thus, it is equivalent to the one posited by the
objective function of the social planner in Boucekkine et al. (2013).\ On
the one hand, this is re-assuring, as (together with the convergence results
in Section 3.2) it shows that the two approaches lead to the same
conclusions. However, one may wonder whether the analysis extends to a less
restrictive sufficiency condition. We show that this is indeed the case; in
fact, the following result holds:

\begin{theorem}
\label{thm:main} Let $\Omega =\mathbb{T}\times \mathbb{R}^{+}$. Assume the
functions $A:\mathbb{R}^{+}\rightarrow \mathbb{R}^{+}$, $k_{0}:\mathbb{T}%
\rightarrow \mathbb{R}^{+}\cup \{0\}$ and $c:\mathbb{T}\times \mathbb{R}%
^{+}\rightarrow \mathbb{R}^{+}\cup \{0\}$ are continuous in their respective
domains. Assume, moreover, that 
\begin{equation}
k_{0}(x)\geq \int_{0}^{t}e^{-\int_{0}^{s}A(z)\,dz}\max_{x\in \mathbb{T}%
}c(x,s)\,ds\quad \forall (x,t)\in \Omega \,.  \label{eq:Ak0c}
\end{equation}

Then the Problem~\ref{Prob1} admits a unique non-negative classical solution.
\end{theorem}

\label{rem:cases} Theorem~\ref{thm:main} only requires that the consumption
function is continuous and that the present discounted value of the flow of
maximum consumption (on the entire space) does not exceed the initial
capital stock at any point of the space. Note that the discounting is done
using a (time-varying) technology parameter.

The economic interpretation of this condition is as follows. Note that (\ref%
{eq:Ak0c}) allows for spatial inequality in consumption; it just imposes the
upper bound on the (present discounted value of the) highest values of this
consumption. Moreover, given the discounting, it allows for an increasing
spatial inequality in consumption over time. The upper bound imposed depends
on the initial distribution of the capital stock, and in particular, the
condition (\ref{eq:Ak0c}) is most stringent for the lowest initial capital
stock on the circle. In other words, it is more difficult to satisfy this
condition when, ceteris paribus, the initial spatial inequality in capital
stock is higher.\footnote{%
The conditions~\eqref{eq:Ak0c1} and~\eqref{eq:Ak0c} hold for all positive $t$%
; hence, they imply that the integral on the right hand side converges at
the limit.\ This implies 
\begin{equation*}
\lim_{t\rightarrow \infty }\left[ e^{-\int_{0}^{t}A(s)\,ds}c(x,t)\right]
=0\quad \forall x\in \mathbb{T}\,,
\end{equation*}%
and 
\begin{equation*}
\lim_{t\rightarrow \infty }\left[ e^{-\int_{0}^{t}A(s)\,ds}\max_{x\in 
\mathbb{T}}c(x,t)\right] =0\,,
\end{equation*}%
which constrains the consumption function to grow over time at a slightly
lower rate than the time-varying technology parameter.}

\subsection{Proof of the main results}

\label{ssec:proof}

Let's start by observing that we can simplify Eq.~\eqref{eq:PDEgen} by
removing the term $A(t)k(x,t)$ using the following Lemma (whose proof is
presented in the Appendix):

\noindent \textbf{Lemma 1 (Equivalent solutions)} \emph{Let $k(x,t)$ and $%
h(x,t)$ be two positive functions defined in $\Omega $ and related to each
other by 
\begin{equation}
h(x,t)=e^{-\int_{0}^{t}A(s)\,ds}k(x,t)\,.
\end{equation}%
Then $k(x,t)$ is a positive solution of~\eqref{eq:PDEgen} with initial
condition~\eqref{eq:IVC} if and only if $h(x,t)$ is a positive solution of 
\begin{equation}
h_{t}=h_{xx}-\gamma (x,t)\quad \forall (x,t)\in \Omega \,,
\label{eq:PDEgen2text}
\end{equation}%
with the same initial condition~\eqref{eq:IVC}, where $\gamma
(x,t)=e^{-\int_{0}^{t}A(s)\,ds}c(x,t)$.}

The proofs of Theorems \ref{thm:main1} and~\ref{thm:main} are constructed
with following steps.\footnote{%
For the sake of clarity, all the proofs except those concerning the
non-negativity are relegated to the Appendix.} We first find a formal
solution, i.e. a Fourier series that, order by order, solves~%
\eqref{eq:PDEgen}. 

\begin{proposition}[Formal solution]
\label{prop:formal} Let us define, for any positive integer $n$, $\lambda
_{n}=n^{2}$ and, for $(x,y,t)\in \mathbb{T}^{2}\times \mathbb{R}^{+}$, the
Green's function 
\begin{equation}
G(x,y,t)=\sum_{n\geq 0}e^{-\lambda _{n}t}\cos \left[ n\left( x-y\right) %
\right] \,.
\end{equation}

Then the function $h(x,t)$ given by 
\begin{equation}
h(x,t)=\frac{1}{\pi }\int_{-\pi }^{\pi }G(x,y,t)k_{0}(y)\,dy-\frac{1}{\pi }%
\int_{0}^{t}\,ds\int_{-\pi }^{\pi }G(x,y,t-s)\gamma (y,s)\,dy\,,
\end{equation}%
is a formal solution of Eq.~\eqref{eq:PDEgen2text}.
\end{proposition}

The second step is to prove that the solution provided by the previous
proposition is actually a classical solution.

\begin{proposition}[Classical solution]
\label{prop:class} Under the above assumptions the function $G(x,y,t)$,
respectively $h(x,t)$, is continuous in $\mathbb{T}^{2}\times \lbrack
0,+\infty )$, respectively $\mathbb{T}\times \lbrack 0,+\infty )$, and twice
differentiable in $\mathbb{T}^{2}\times (0,+\infty )$, respectively $\mathbb{%
T}\times (0,+\infty )$.
\end{proposition}

The third steps is to prove the uniqueness of the classical solution:

\begin{proposition}[Uniqueness]
\label{prop:uniq} The classical solution of the Eq.~\eqref{eq:PDEgen2text}
with initial condition $k(x,t)=k_{0}(x)$ is unique.
\end{proposition}

Considering the above-mentioned results and using Lemma~\ref{lem:equiv}, we
can conclude that 
\begin{eqnarray}
k(x,t) &=&h(x,t)e^{\int_{0}^{t}A(s)\,ds}  \label{eq:solution} \\
:= &&e^{\int_{0}^{t}A(s)\,ds}\left[ \frac{1}{\pi }\int_{-\pi }^{\pi
}G(x,y,t)k_{0}(y)\,dy-\frac{1}{\pi }\int_{0}^{t}\,ds\int_{-\pi }^{\pi
}G(x,y,t-s)\gamma (y,s)\,dy\right] \,,  \notag
\end{eqnarray}%
is the unique classical solution of~\eqref{eq:PDEgen} with initial condition 
$k_{0}(x)$.

The last step for proving Theorem~\ref{thm:main1} is to show that the
solution given by~\eqref{eq:solution} is non-negative for all $(x,t)\in
\Omega $, that is the function describing the evolution of the stock of
capital in space and time $k(x,t)$ is everywhere and always non-negative.
For this, we need a preliminary result (Lemma~\ref{lem:posit}) whose proof
can be found in the Appendix.

\begin{proposition}[Non-negativity]
\label{prop:posit1} Let $h(x,t)$ be the classical solution of the Eq.~%
\eqref{eq:PDEgen2text} in $\Omega $ with boundary condition $%
h(x,0)=k_{0}(x)\geq 0$ for all $x\in \mathbb{T}$ and assume $\gamma
(x,t)\geq 0$ for all $(x,t)\in \Omega $. Under the hypotheses 
\begin{equation*}
\gamma _{xx}(x,t)\leq 0\quad \forall (x,t)\in \Omega \,,
\end{equation*}%
and 
\begin{equation*}
k_{0}(x)\geq \int_{0}^{t}\gamma (x,s)\,ds\quad \forall (x,t)\in \Omega \,,
\end{equation*}%
we have 
\begin{equation}
h(x,t)\geq 0\quad \forall (x,t)\in \Omega \,.  \label{eq:asspos2}
\end{equation}
\end{proposition}

\proof
Let $T>0$, $\epsilon>0$ and let us define the auxiliary function 
\begin{equation*}
v(x,t)=h(x,t)+\epsilon t+\int_0^t \gamma(x,s)\, ds\, ,
\end{equation*}
for all $(x,t)\in\mathbb{T}\times [0,T]$.

A straightforward computation gives: 
\begin{equation*}
v_{t}-v_{xx}=h_{t}-h_{xx}+\epsilon +\gamma (x,t)-\int_{0}^{t}\gamma
_{xx}(x,s)\,ds=\epsilon -\int_{0}^{t}\gamma _{xx}(x,s)\,ds\geq \epsilon >0,
\end{equation*}%
where we use the fact that $h_{t}-h_{xx}=-\gamma (x,t)$ and the assumption $%
\gamma _{xx}(x,s)\leq 0$. We can thus apply the Lemma~\ref{lem:posit} to $v$
and conclude that it attains its minimum at some $(a,\tau )\in \mathbb{T}%
\times \{0\}$.

Hence, we have that for all $(x,t)\in \mathbb{T}\times \lbrack 0,T]$ 
\begin{equation*}
h(x,t)+\epsilon t+\int_{0}^{t}\gamma (x,s)\,ds=v(x,t)\geq
v(x,0)=h(x,0)=k_{0}(x)\,,
\end{equation*}%
thus 
\begin{equation*}
h(x,t)+\epsilon t\geq k_{0}(x)-\int_{0}^{t}\gamma (x,s)\,ds\geq 0\,,
\end{equation*}%
where the rightmost inequality holds because of the assumption on $k_{0}$.
We can finally pass to the limit $\epsilon \rightarrow 0$ and conclude that 
\begin{equation*}
h(x,t)\geq 0\quad (x,t)\in \mathbb{T}\times \lbrack 0,T]\,.
\end{equation*}%
The arbitrariness of $T$ completes the proof. \endproof

Because the function $k(x,t)$ has the same sign that $h(x,t)$, we conclude
that $k(x,t)$ is also non-negative in $\Omega $ and this concludes the proof
of Theorem~\ref{thm:main1}.

The proof of Theorem~\ref{thm:main} can be achieved in a very similar way;
the only difference lies on the way we prove the non-negativity of the
solution~\eqref{eq:solution} under the assumptions of Theorem~\ref{thm:main}.

\begin{proposition}[Non-negativity]
\label{prop:posit} Let $h(x,t)$ be the classical solution of the Eq.~%
\eqref{eq:PDEgen2text} in $\Omega $ with boundary condition $%
h(x,0)=k_{0}(x)\geq 0$ for all $x\in \mathbb{T}$ and assume $\gamma
(x,t)\geq 0$ for all $(x,t)\in \Omega $. Under the hypothesis 
\begin{equation*}
k_{0}(x)\geq \int_{0}^{t}\max_{x\in \mathbb{T}}\gamma (x,s)\,ds\quad \forall
(x,t)\in \Omega \,,
\end{equation*}%
we have 
\begin{equation}
h(x,t)\geq 0\quad \forall (x,t)\in \Omega \,.  \label{eq:asspos}
\end{equation}
\end{proposition}

\proof
Let $T>0$, $\epsilon>0$ and let us define the auxiliary function 
\begin{equation*}
v(x,t)=h(x,t)+\epsilon t+\int_0^t \max_{x\in\mathbb{T}}\gamma(x,s)\, ds\, ,
\end{equation*}
for all $(x,t)\in\mathbb{T}\times [0,T]$.

A direct computation provides $v_{t}-v_{xx}\geq 0$ and, thus, by Lemma~\ref%
{lem:posit}, $v$ attains its minimum at some $(a,\tau )\in \mathbb{T}\times
\{0\}$. Then, following an argument similar to the one used in the previous
Proposition we conclude that $h(x,t)+\epsilon t\geq
k_{0}(x)-\int_{0}^{t}\max_{x\in \mathbb{T}}\gamma (x,s)\,ds\geq 0$. Passing
to the limit $\epsilon \rightarrow 0$, and using the arbitrariness of $T$,
we get the result. \endproof

Once again, because $k(x,t)$ and $h(x,t)$ differ by a positive function, we
can conclude that $k(x,t)$ is non-negative in $\Omega $.

Our results above show that if the initial spatial inequality in capital is
not too stark and that the consumption does not differ too much across space
(so as to respect (\ref{eq:Ak0c})), a given consumption function uniquely
determines the spatial growth process. However, the natural question remains
whether such dynamics (always) leads to convergence of capital stock across
space over time. We address this question below.

\subsection{Asymptotic behavior and convergence}

\label{ssec:asymptbehav}

The aim of this sub-section is to study the asymptotic behavior of the
classical solution $k(x,t)$ for large $t$. Our analysis relies on the
behavior of the Green's function $G(x,y,t)$ for large $t$ (that is fully
characterized by Lemma~\ref{lemma:Gtlarge} in the Appendix).

Let us first consider the case of time-independent technology, i.e. $%
A(t)=A_{0}$ for all $t$. Our main result is the following

\begin{proposition}
\label{prop:ktlarge} Let (\ref{eq:Ak0c}) hold and $k(x,t)$ be the classical
non-negative solution of~\eqref{eq:PDEgen} with initial condition $%
k(x,0)=k_{0}(x)$ and $A(t)=A_{0}\in \mathbb{R}^{+}$. Then, assuming $%
\widetilde{c}(t)$ to be bounded, we have 
\begin{equation}
\lim_{t\rightarrow \infty }\left[ k(x,t)e^{-A_{0}t}\right] =\widetilde{k}%
_{0}-\int_{0}^{\infty }e^{-A_{0}s}\widetilde{c}(s)\,ds\,  \label{eq:klarget}
\end{equation}%
uniformly in $\mathbb{T}$, where 
\begin{equation*}
\widetilde{k}_{0}=\frac{1}{\pi }\int_{-\pi }^{\pi }k_{0}(x)\,dx\quad \text{%
and}\quad \widetilde{c}(t)=\frac{1}{\pi }\int_{-\pi }^{\pi }c(x,t)\,dx\,.
\end{equation*}
\end{proposition}

\proof From the explicit form for the classical non-negative solution $k(x,t)
$ in the case $A(t)=A_{0}$ we get: 
\begin{equation*}
k(x,t)e^{-A_{0}t}=\frac{1}{\pi }\int_{-\pi }^{\pi }G(x,y,t)k_{0}(y)\,dy-%
\frac{1}{\pi }\int_{0}^{t}\,ds\int_{-\pi }^{\pi
}G(x,y,t-s)e^{-A_{0}s}c(y,s)\,dy\,,
\end{equation*}%
that can be rewritten as 
\begin{eqnarray*}
k(x,t)e^{-A_{0}t} &=&\frac{1}{\pi }\int_{-\pi }^{\pi }\left[ G(x,y,t)-1%
\right] k_{0}(y)\,dy+\widetilde{k}_{0} \\
&-&\frac{1}{\pi }\int_{0}^{t}\,ds\int_{-\pi }^{\pi }\left[ G(x,y,t-s)-1%
\right] e^{-A_{0}s}c(y,s)\,dy-\int_{0}^{t}e^{-A_{0}s}\widetilde{c}(s)\,ds\,.
\end{eqnarray*}

Using Lemma~\ref{lemma:Gtlarge} the first term on the top line becomes, at
the limit $t\rightarrow \infty $: 
\begin{equation*}
\lim_{t\rightarrow \infty }\left[ \frac{1}{\pi }\int_{-\pi }^{\pi }\left[
G(x,y,t)-1\right] k_{0}(y)\,dy\right] =0\,.
\end{equation*}

The remaining terms can be handled as follows: 
\begin{gather*}
\left\vert \frac{1}{\pi }\int_{0}^{t}\,ds\int_{-\pi }^{\pi }\left[
G(x,y,t-s)-1\right] e^{-A_{0}s}c(y,s)\,dy\right\vert \leq \frac{1}{\pi }%
\int_{0}^{t}\,ds\int_{-\pi }^{\pi }\sum_{n\geq
1}e^{-n^{2}t}e^{(n^{2}-A_{0})s}c(y,s)\,dy \\
\leq \max_{s\leq t}\widetilde{c}(s)\sum_{n\geq 1}e^{-n^{2}t}\frac{1}{%
n^{2}-A_{0}}\left( e^{(n^{2}-A_{0})t}-1\right) \\
\leq \max_{s\leq t}\widetilde{c}(s)\left( e^{-A_{0}t}\sum_{n\geq 1}\frac{1}{%
n^{2}-A_{0}}-\sum_{n\geq 1}\frac{e^{-n^{2}t}}{n^{2}-A_{0}}\right) \,.
\end{gather*}%
The series are convergent and their sums vanish in the limit. Hence, 
\begin{equation*}
\lim_{t\rightarrow \infty }\left[ \frac{1}{\pi }\int_{0}^{t}\,ds\int_{-\pi
}^{\pi }\left[ G(x,y,t-s)-1\right] e^{-A_{0}s}c(y,s)\,dy\right] =0\,,
\end{equation*}%
which concludes the proof. \endproof

Let us now consider the case of a time-dependent technology parameter. One
can prove a similar result under the additional minor assumption that the
technology parameter is always strictly positive, $A(t)>0$.

\begin{proposition}
\label{prop:ktlarge1} Let condition (\ref{eq:Ak0c}) hold and $k(x,t)$ be the
classical positive solution of~\eqref{eq:PDEgen} with initial condition $%
k(x,0)=k_{0}(x)$, $A:\mathbb{R}^{+}\rightarrow \mathbb{R}^{+}$, and there
exists $A_{+}>0$ such that $A(t)\geq A_{+}$ for all $t\geq 0$. Then,
assuming $\widetilde{c}(t)$ to be bounded, we get%
\begin{equation}
\lim_{t\rightarrow \infty }\left[ k(x,t)e^{-\int_{0}^{t}A(s)\,ds}\right] =%
\widetilde{k}_{0}-\int_{0}^{\infty }e^{-\int_{0}^{s}A(z)\,dz}\widetilde{c}%
(s)\,ds\,.  \label{eq:klarget1}
\end{equation}%
uniformly in $\mathbb{T}$, where%
\begin{equation*}
\widetilde{k}_{0}=\frac{1}{\pi }\int_{-\pi }^{\pi }k_{0}(x)\,dx\quad \text{%
and}\quad \widetilde{c}(t)=\frac{1}{\pi }\int_{-\pi }^{\pi }c(x,t)\,dx\,.
\end{equation*}
\end{proposition}

\proof The claim can be proven along the same lines of the previous
Proposition, observing that we have the following bound: 
\begin{equation*}
e^{-\int_{0}^{t}A(s)\,ds}\leq e^{-A_{+}t}\quad \forall \,t\geq 0\,.
\end{equation*}%
\endproof

We have shown that any optimal growth dynamics that satisfies the condition (%
\ref{eq:Ak0c}) leads to spatial convergence in capital over time.
Intuitively, the capital diffusion based on spatial trade between more
capital-abundant locations and their less capital-abundant neighbors, as
described by the trade balance relation $\tau (x,t)=-k_{xx}(x,t)$,
guarantees, for a given consumption profile that respects (\ref{eq:Ak0c}),
that the initial spatial inequality in capital disappears over time.
Crucially, this does not depend on a specific objective function of the
social planner. This is important because setting the spatial weights in
such objective function as being equal for every location is somewhat
arbitrary and might not respect certain welfare criteria; our findings above
imply that this equality-of-weights assumption can be relaxed.

\section{Conclusion}

We have shown that the asymptotic convergence in a stylized spatial AK
growth model does not depend on restrictive assumptions about the objective
function of the social planner. We have also generalized this finding,
allowing for the time-varying technology parameter, and provided an explicit
solution for the dynamics of spatial distribution of the capital stock.

Two directions for future research look particularly promising. First, the
technology might depend not only on time but also differ in space, as
suggested, for instance, by Quah (2002).\ This conjecture has been confirmed
empirically. For instance, large spatial productivity differences have been
documented by Acemoglu and Dell (2010) for Latin America, which also show
that within-country differences are much larger than the between-country
ones, and suggest that these differences are shaped by institutional
features (in particular, by the distribution of political power locally). An
alternative explanation is that agglomeration externalities make firms more
productive, as has been shown by Combes et al. (2012). This probably has to
do with the space-varying innovation incentives of firms, as discussed and
modelled by Desmet and Rossi-Hansberg (2012, 2014).\ This calls for an
analysis of the growth dynamics with space- and time-varying technological
parameter, along the lines of this paper.\footnote{%
Notably, Comin et al. (2013) document the fundamental importance of the
geographic distance for technology diffusion, which shows why an explicit
modelling of the spatial dimension is crucial.}

Second, the spatial dynamics of the capital stock might not always be
described appropriately by a diffusion process.\ Empirically, Desmet and
Rossi-Hansberg (2009) show that diffusive dynamics applies well to
established sectors (such as manufacturing in 1970-2000s); however, for
younger sectors (e.g., manufacturing at the beginning of the 20th century or
retail and financial sectors in 1970-2000s), the dynamics is (at least
locally) agglomerative.\ This requires a modelling of the equation of motion
of capital richer than the one presented in this paper. Ideally, such a
model would also capture the transition process from a non-diffusive to a
diffusive dynamics.

\appendix

\section{Technical proofs}

\label{sec:appproofx}

\begin{lemma}
\label{lem:equiv} Let $k(x,t)$ and $h(x,t)$ be two positive functions
defined in $\Omega $ and related to each other by%
\begin{equation}
h(x,t)=e^{-\int_{0}^{t}A(s)\,ds}k(x,t)\,.  \label{eq:equiv}
\end{equation}%
Then $k(x,t)$ is a positive solution of~\eqref{eq:PDEgen} with initial
condition~\eqref{eq:IVC} if and only if $h(x,t)$ is a positive solution of 
\begin{equation}
h_{t}=h_{xx}-\gamma (x,t)\quad \forall (x,t)\in \Omega \,,
\label{eq:PDEgen2b}
\end{equation}%
where $\gamma (x,t)=e^{-\int_{0}^{t}A(s)\,ds}c(x,t)$, with the \emph{initial
condition}~\eqref{eq:IVC}.
\end{lemma}

\proof The proof is obtained by direct computation. In fact, for all $%
(x,t)\in \Omega $ we get: 
\begin{eqnarray*}
h_{t}-h_{xx}
&=&-A(t)h(x,t)+e^{-\int_{0}^{t}A(s)\,ds}k_{t}(x,t)-e^{-\int_{0}^{t}A(s)%
\,ds}k_{xx}(x,t) \\
&=&-A(t)h(x,t)+e^{-\int_{0}^{t}A(s)\,ds}\left[ A(t)k(x,t)-c(x,t)\right]
=-e^{-\int_{0}^{t}A(s)\,ds}c(x,t)=-\gamma (x,t)\,.
\end{eqnarray*}%
Finally 
\begin{equation*}
h(x,0)=k(x,0)=k_{0}(x)\,.
\end{equation*}%
\endproof

\textbf{Proposition 1 (Formal solution).} \textit{Let us define, for any
positive integer }$n$\textit{, }$\lambda _{n}=n^{2}$\textit{\ and let us
define for }$(x,y,t)\in T^{2}\times R^{+}$\textit{, the Green's function }%
\begin{equation}
G(x,y,t)=\sum_{n\geq 0}e^{-\lambda _{n}t}\cos \left[ n\left( x-y\right) %
\right] \,.  \label{eq:kernel}
\end{equation}

\textit{Then the function }$h(x,t)$\textit{\ given by }%
\begin{equation}
h(x,t)=\frac{1}{\pi }\int_{-\pi }^{\pi }G(x,y,t)k_{0}(y)\,dy-\frac{1}{\pi }%
\int_{0}^{t}\,ds\int_{-\pi }^{\pi }G(x,y,t-s)\gamma (y,s)\,dy\,,
\label{eq:formsol}
\end{equation}%
\textit{is a formal solution of Eq.~\eqref{eq:PDEgen2b}.}

\proof Let us observe that formal Green's function satisfies for all $t>0$
and $(x,y)\in \mathbb{T}^{2}$ the following homogeneous PDE 
\begin{equation}
\partial _{t}G(x,y,t)=\partial _{x}^{2}G(x,y,t)\,,  \label{eq:kerneleq}
\end{equation}%
and moreover, 
\begin{equation}
\lim_{t\rightarrow 0}G(x,y,t)=\delta (x-y)\,,  \label{eq:kerneldelta}
\end{equation}%
where $\delta (x-y)$ is the Dirac delta-function.

Then, to prove our claim, it is sufficient to plug $h(x,t)$ given by~%
\eqref{eq:formsol} into Eq.~\eqref{eq:PDEgen2b} and by exchanging the
derivatives with the integrals one easily finds: 
\begin{eqnarray*}
h_{t}-h_{xx} &=&\frac{1}{\pi }\int_{-\pi }^{\pi }\left[ \partial
_{t}G(x,y,t)-\partial _{x}^{2}G(x,y,t)\right] k_{0}(y)\,dy \\
&-&\frac{1}{\pi }\int_{0}^{t}\,ds\int_{-\pi }^{\pi }\left[ \partial
_{t}G(x,y,t-s)-\partial _{x}^{2}G(x,y,t-s)\right] \gamma (y,s)\,dy\,, \\
&-&\lim_{t\rightarrow s}\frac{1}{\pi }\int_{-\pi }^{\pi }G(x,y,t-s)\gamma
(y,s)\,dy\,,
\end{eqnarray*}

Using Eq.~\eqref{eq:kerneleq} the first two terms on the right-hand side
vanish, while using~\eqref{eq:kerneldelta} the remaining term reduces to $%
\gamma (x,t)$. We can thus conclude that: 
\begin{equation*}
h_{t}-h_{xx}=-\gamma (x,t)\,,
\end{equation*}%
hence $h(x,t)$ solves~\eqref{eq:PDEgen2b}.

To show that $h(x,t)$ satisfies the initial condition~\eqref{eq:IVC}, it is
enough to pass to the limit $t\rightarrow 0$ in the definition~%
\eqref{eq:formsol} and to make use of~\eqref{eq:kerneldelta}. \endproof

\textbf{Proposition 2 (Classical solution). }\textit{Under the above
assumptions, the function }$G(x,y,t)$\textit{, respectively }$h(x,t)$\textit{%
, is continuous in }$T^{2}\times \lbrack 0,+\infty )$\textit{, respectively }%
$T\times \lbrack 0,+\infty )$\textit{, and twice differentiable in }$%
T^{2}\times (0,+\infty )$\textit{, respectively }$T\times (0,+\infty )$%
\textit{.}

\proof The claim is proven by showing that $G(x,y,t)$ is the uniform limit
of smooth functions on $\Omega ,$ and thus it is itself a smooth function.

We start by rewriting~\eqref{eq:kernel} as: 
\begin{equation*}
G(x,y,t)=1+\sum_{n\geq 1}e^{-\lambda _{n}t}\cos \left[ n\left( x-y\right) %
\right] \,,
\end{equation*}%
hence, after observing that for all $t>0$ and $(x,y)\in \mathbb{T}^{2}$, one
has: 
\begin{equation*}
\lvert e^{-\lambda _{n}t}\cos \left[ n\left( x-y\right) \right] \rvert \leq
e^{-n^{2}t}\,,
\end{equation*}%
and we can conclude that 
\begin{equation*}
\Big\lvert\sum_{n\geq 1}e^{-\lambda _{n}t}\cos \left[ n\left( x-y\right) %
\right] \Big\rvert\leq e^{-t}\sum_{m\geq 0}e^{-2mt}=\frac{e^{-t}}{1-e^{-2t}}%
\,,
\end{equation*}%
which proves the norm convergence of the sum.\ This, in turn, implies the
uniform convergence and thus the smoothness of $G(x,y,t)$.

The claim for $h(x,t)$ follows easily from its definition and is thus
skipped. \endproof

\textbf{Proposition 3 (Uniqueness).} \textit{The classical solution of the
Eq.~\eqref{eq:PDEgen2b} with initial condition }$k(x,t)=k_{0}(x)$\textit{\
is unique.}

\proof Let us assume, on the contrary, the existence of two distinct
classical solutions $h_{1}(x,t)$ and $h_{2}(x,t)$ of the Eq.~%
\eqref{eq:PDEgen2b} with initial conditions $h_{i}(x,0)=k_{0}(x)$ for $i=1,2$%
. Then, the function $f(x,t)=h_{1}(x,t)-h_{2}(x,t)$ solves the equation 
\begin{equation*}
f_{t}=f_{xx}\quad \forall (x,t)\in \Omega \,,
\end{equation*}%
with initial condition $f(x,0)=0$. However, then, using the Green's function 
$G(x,y,t)$ defined above, one easily concludes that $f(x,t)$ identically
vanishes on $\Omega $ and thus $h_{1}\equiv h_{2}$.

\begin{lemma}
\label{lem:posit} Let $T>0$ and let $v$ be a smooth function satisfying the
inequality 
\begin{equation*}
v_{t}-v_{xx}>0\quad \forall (x,t)\in \mathbb{T}\times \lbrack 0,T]\,.
\end{equation*}%
Then, $v$ attains its minimum at some $(a,\tau )\in \mathbb{T}\times \{0\}$.
\end{lemma}

\proof Because of the smoothness assumption, $v$ must attain its minimum
somewhere on the compact set $\mathbb{T}\times \lbrack 0,T]$. Let's assume,
by contradiction, the minimum $(a,\tau )$ to lie in $\mathbb{T}\times (0,T]$%
. Then, by elementary calculus we have~\footnote{%
Actually one can have $v_{t}(a,\tau )<0$ only if $\tau <T$.} 
\begin{equation*}
v_{t}(a,\tau )\leq 0\quad \text{and}\quad v_{xx}(a,\tau )\geq 0\,,
\end{equation*}%
from which we straightforwardly calculate: 
\begin{equation*}
v_{t}(a,\tau )-v_{xx}(a,\tau )\leq 0\,,
\end{equation*}%
contradicting the hypothesis. Hence, we must have $(a,\tau )\in \mathbb{T}%
\times \{0\}$. \endproof

\begin{lemma}
\label{lemma:Gtlarge} Let $G(x,y,t)$ be the Green's function defined by~%
\eqref{eq:kernel}; then 
\begin{equation}
\lim_{t\rightarrow \infty }\lvert G(x,y,t)-1\rvert =0\,,  \label{eq:Glarget}
\end{equation}%
uniformly for $(x,y)\in \mathbb{T}^{2}$.
\end{lemma}

\proof Using the definition, we can write 
\begin{equation*}
G(x,y,t)-1=\sum_{n\geq 1}e^{-\lambda _{n}t}\cos \left[ n\left( x-y\right) %
\right] \,,
\end{equation*}%
from which we get 
\begin{equation*}
\lvert G(x,y,t)-1\rvert \leq \sum_{n\geq 1}e^{-n^{2}t}\leq \frac{e^{-t}}{%
1-e^{-2t}}\,.
\end{equation*}%
Hence Eq.~\eqref{eq:Glarget} follows directly. \endproof


\begin{thebibliography}{99}
\bibitem{acemoglu2009} Acemoglu, D. 2008.\ \textit{Introduction to Modern
Economic Growth}.\ Princeton, NJ: Princeton University Press.

\bibitem{ad} Acemoglu, D., Dell, M. 2010. Productivity differences between
and within countries. \textit{American Economic Journal: Macroeconomics} 
\textbf{2}: 169-188.

\bibitem{boucekkine2013} Boucekkine, R., Camacho, C., Fabbri, G. 2013.
Spatial dynamics and convergence: The spatial AK model. \textit{Journal of
Economic Theory} \textbf{148}: 2719--2736.

\bibitem{boucekkine2009} Boucekkine, R., Camacho, C., Zou, B. 2009. Bridging
the gap between growth theory and the new economic geography: the spatial
Ramsey model. \textit{Macroeconomic Dynamics} \textbf{13}: 20-45.

\bibitem{brito2004} Brito, P. 2004. The dynamics of growth and distribution
in a spatially heterogeneous world. Department of Economics, ISEG, Working
Paper WP13/2004/DE/UECE.

\bibitem{combes} Combes, P.-Ph., Duranton, G., Gobillon, L., Puga, D., Roux,
S. 2012. The productivity advantages of large cities: Distinguishing
agglomeration from firm selection. \textit{Econometrica} \textbf{80}:
2543-2594.

\bibitem{cominetal} Comin, D., Dmitriev, M., Rossi-Hansberg, E. 2013. The
spatial diffusion of technology. Working paper, Harvard University.

\bibitem{drh2009} Desmet, K., Rossi-Hansberg, E. 2009. Spatial growth and
industry age. \textit{Journal of Economic Theory} \textbf{144}: 2477--2502.

\bibitem{drh2010} Desmet, K., Rossi-Hansberg, E. 2010. On spatial dynamics. 
\textit{Journal of Regional Science} \textbf{50}: 43-63.

\bibitem{drh2012} Desmet, K., Rossi-Hansberg, E. 2012. Innovation in space. 
\textit{American Economic Review Papers and Proceedings }\textbf{102}:
447-452

\bibitem{drh2013} Desmet, K., Rossi-Hansberg, E. 2014. Spatial development. 
\textit{American Economic Review} (forthcoming)\textbf{.}

\bibitem{isard1979} Isard, W., Liossatos, P. 1979. Spatial Dynamics and
Optimal Space-Time Development. Amsterdam: North-Holland.

\bibitem{lucas1988} Lucas, R. 1988.\ On the mechanics of economic
development.\ \textit{Journal of Monetary\ Economics} \textbf{22}: 3-42.

\bibitem{krugman1997} Krugman, P. 1997. \textit{Development, Geography, and
Economic\ Theory}. Cambridge, MA: MIT\ Press.

\bibitem{romer1986} Romer, P. 1986. Increasing returns and long-run growth. 
\textit{Journal of\ Political Economy} \textbf{94}: 1002-1037.

\bibitem{rebelo1991} Rebelo, S. 1991. Long-run policy analysis and long-run
growth. \textit{Journal of Political Economy} \textbf{99}: 500-521.

\bibitem{quah2002} Quah, D. 2002. Spatial Agglomeration Dynamics. \textit{%
American Economic Review }\textbf{92}: 247-252.
\end{thebibliography}
\end{document}